\documentclass[12pt]{article}

\usepackage{cite}
\usepackage{epsfig}
\usepackage{amsmath}
\usepackage{array}

\def\mathswitch#1{\relax\ifmmode#1\else$#1$\fi}
\def\mathswitchr#1{\relax\ifmmode{\mathrm{#1}}\else$\mathrm{#1}$\fi}
\newcommand{\PW}{\mathswitchr W}
\newcommand{\PZ}{\mathswitchr Z}
\newcommand{\PH}{\mathswitchr H}

\newcommand{\Pb}{\mathswitchr b}
\newcommand{\Pt}{\mathswitchr t}

\newcommand{\Pf}{\mathswitchr f}
\newcommand{\MW}{\mathswitch {M_\PW}}
\newcommand{\MZ}{\mathswitch {M_\PZ}}
\newcommand{\GZ}{\mathswitch {\Gamma_\PZ}}
\newcommand{\MH}{\mathswitch {M_\PH}}

\newcommand{\mb}{\mathswitch {m_\Pb}}
\newcommand{\mt}{\mathswitch {m_\Pt}}
\newcommand{\mf}{m_f}
\newcommand{\scrs}{\scriptscriptstyle}
\newcommand{\sw}{\mathswitch {s_{\scrs\PW}}}

\newcommand{\mw}{\mathswitch {\overline{M}_\PW}}
\newcommand{\mz}{\mathswitch {\overline{M}_\PZ}}

\newcommand{\gz}{\mathswitch {\overline{\Gamma}_\PZ}}
\newcommand{\as}{\alpha_{\mathrm s}}

\newcommand{\gev}{\,\, \mathrm{GeV}}
\newcommand{\mev}{\,\, \mathrm{MeV}}

\newcommand{\SLASH}[2]{\makebox[#2ex][l]{$#1$}/}

\newcommand{\pslash}{\SLASH{p}{.2}}

\newcommand{\anc}{\rule{0mm}{0mm}}

\newcommand{\OO}{{\mathcal O}}

\newcommand{\mycaption}[1]{\caption{\sl #1}}

\makeatletter
\def\section{\@startsection {section}{1}{\z@}{+3.0ex plus +1ex minus
  +.2ex}{2.3ex plus .2ex}{\large\bf\boldmath}}
\def\subsection{\@startsection{subsection}{2}{\z@}{+2.5ex plus +1ex
minus +.2ex}{1.5ex plus .2ex}{\normalsize\bf\boldmath}}
\def\subsubsection{\@startsection{subsubsection}{3}{\z@}{+3.25ex plus
 +1ex minus +.2ex}{1.5ex plus .2ex}{\normalsize\it}}

\graphicspath{{.}{plots/}}

\begin{document}
\thispagestyle{empty}

\def\thefootnote{\fnsymbol{footnote}}

\begin{flushright}
\end{flushright}

\vspace{1cm}

\begin{center}

{\Large {\bf Higher-order electroweak corrections to the\\[1ex]
partial widths and branching ratios of the \boldmath $Z$ boson}}
\\[3.5em]
{\large
Ayres~Freitas
}

\vspace*{1cm}

{\sl
Pittsburgh Particle-physics Astro-physics \& Cosmology Center
(PITT-PACC),\\ Department of Physics \& Astronomy, University of Pittsburgh,
Pittsburgh, PA 15260, USA
}

\end{center}

\vspace*{2.5cm}

\begin{abstract}

Recently, the calculation of fermionic electroweak two-loop corrections to the
total width of the $Z$ boson and hadronic $Z$-peak cross-section in the Standard
Model has been presented, where ``fermionic'' refers to diagrams with closed
fermion loops. Here, these results are complemented by presenting contributions
of the same order for the $Z$-boson partial widths, which are the last missing
pieces for a complete description of $Z$-pole physics at the fermionic two-loop
order. The definition of the relevant observables and the calculational
techniques are described in detail. Numerical results are presented conveniently
in terms of simple parametrization formulae. Finally, the remaining
theoretical uncertainties from missing higher-order corrections are analyzed and
found to be small compared to the current experimental errors.
\end{abstract}

\setcounter{page}{0}
\setcounter{footnote}{0}

\newpage

%%%%%%%%%%%%%%%%%%%%%%%%%%%%%%%%%%%%%%%%%%%%%%%%%%%%%%%%%%%%%%

\section{Introduction}

Electroweak precision observables (EWPOs) play a crucial role in testing the
Standard Model (SM) at the quantum level and constraining physics beyond the SM.
Some of the most important quantities in this context are the masses and widths
of the $W$ and $Z$ bosons and the $Z$-boson couplings. The latter have been
measured through the cross-section and polarization and angular asymmetries of
the process $e^+e^- \to (Z) \to f\bar{f}$ at LEP1 and SLC, see
$e.\,g.$~Ref.~\cite{lep1}. Here $f$ stands for any SM lepton or quark, except
the top quark, and the symbol $(Z)$ in brackets indicates $s$-channel exchange
of an intermediate $Z$ boson, which is
unstable and thus not an asymptotic on-shell state. When computing theoretical
predictions for the $W$ and $Z$ masses, widths and couplings, one has to take
into account loop corrections, which depend on other elements of the SM, such as
the top-quark mass, $\mt$, the Higgs boson mass, $\MH$, and the strong couplings
constant, $\as$. By combining direct measurements of these quantities with EWPOs
in a global fit, one obtains a highly non-trivial and overconstrained test of
the SM. On the other hand, a significant deviation between measurement and SM
fit could be an indication for the presence of new particles in the loop
corrections. For recent examples of such global fits and constraints on new
physics, see $e.\,g.$~Ref.~\cite{gfit,xfit,pdg}.

Owing to the high precision of the experimental measurements, it is mandatory to
include higher-order corrections beyond the one-loop approximation in the theory
calculations. For the SM prediction of the $W$-boson mass, $\MW$, complete
two-loop corrections, of order $\OO(\alpha\as)$ and $\OO(\alpha^2)$, are known
\cite{qcd2,mw,mwlong,mwtot}. In addition, partial three- and four-loop results,
enhanced by powers of $\mt$, have been computed at order $\OO(\alpha\as^2)$
\cite{qcd3}, $\OO(\alpha^2\as\mt^4)$, $\OO(\alpha^3\mt^6)$ \cite{mt6}, and
$\OO(\alpha\as^3\mt^2)$ \cite{qcd4}. The same order of electroweak (EW) and QCD
corrections are available for the leptonic
effective weak mixing angle, $\sin^2 \theta_{\rm eff}^\ell$
\cite{swlept,swlept2}, which describes the ratio of vector and
axial-vector couplings of the $Z$ boson to leptons. The effective weak mixing
angles for quarks have been computed with \emph{fermionic} two-loop corrections,
which stem from diagrams with one or two closed fermion loops \cite{swlept2,swbb},
and the same partial three- and four-loop contributions mentioned above. The
fermionic corrections are enhanced by powers of $\mt$ and the large number of
light fermion flavors, and thus expected to dominate over the \emph{bosonic}
corrections, which correspond to diagrams without closed fermion loop. This
expectation is corroborated by experience from the calculation of $\MW$ and
$\sin^2 \theta_{\rm eff}^\ell$, and thus the theory uncertainty from the missing
bosonic $\OO(\alpha^2)$ corrections is estimated to be relatively small \cite{swlept2,swbb}.

While the effective weak mixing angles are sensitive to the ratio of vector and
axial-vector couplings of the $Z$ boson to fermions, the overall strength of
these couplings can be determined from the measurement of the partial
widths, $\Gamma_f$, for the decay $Z \to f\bar{f}$. However, for the SM
calculation of the $\Gamma_f$, even the fermionic two-loop corrections are not
known, leading to potentially sizeable uncertainties in electroweak precision
tests. The most precise existing result is based on a large-$\mt$ expansion for
the EW two-loop corrections, up to the next-to-leading order
$\OO(\alpha^2\mt^2)$ for final-state leptons and quarks of the
first two generations \cite{ewmt2}, and only up to the leading 
$\OO(\alpha^2\mt^4)$ term for the $Z\to b\bar{b}$ partial width \cite{ewmt4}.
As a first step to improve on this situation, Ref.~\cite{rb} reported on the
calculation of the fermionic two-loop contributions, without any approximation,
to the branching ratio of the $Z$-boson into $b\bar{b}$ and all hadronic final
states, $R_b \equiv \Gamma_b/\Gamma_{\rm had}$.

This article describes the completion of the missing pieces to arrive at a
complete description of $Z$-pole physics at the fermionic two-loop order. For
this purpose, the full fermionic $\OO(\alpha^2)$ corrections to the $Z$-boson
partial widths have been calculated within the SM. Results for two related
quantities, the total $Z$ width, $\GZ$, and the hadronic peak cross-section$,
\sigma^0_{\rm had}$, have been presented recently in Ref.~\cite{short}. In this
paper, details of the calculation are given, and numerical results for the
partial widths, $\Gamma_f$, for all final states $f$ are presented. It is
demonstrated that the inclusion of these new results leads to predictions for
$\Gamma_f$ with a theory uncertainty safely below the experimental error.

The paper is organized as follows. In section~\ref{sc:def}, the definition of
the relevant observables is discussed, with particular attention to gauge
invariance and internal consistency to next-to-next-to-leading order in
perturbation theory. Section~\ref{sc:calc} describes the methods used for the
calculation of the EW two-loop diagrams, which use a combination of
analytical and numerical techniques. Numerical results for the $Z$ partial
widths, as well as for various commonly used branching ratios, are shown in
section~\ref{sc:res}. To make these results available to other researchers,
simple parametrization formulae are provided, which accurately describe the full
calculation within experimentally allowed ranges of the input parameters.
Finally, section~\ref{sc:error} is devoted to a discussion of the remaining
theory error from unknown higher-order corrections.

%%%%%%%%%%%%%%%%%%%%%%%%%%%%%%%%%%%%%%%%%%%%%%%%%%%%%%%%%%%%%%

\section{Definition of the observables}
\label{sc:def}

Since the $Z$ boson is unstable, it cannot be described as an asymptotic state,
and the decay process $Z \to f\bar{f}$ is ill-defined in the usual formalism
of perturbation theory. Instead, for the analysis of $Z$ physics at LEP and SLC,
one needs to consider the process $e^+e^- \to f\bar{f}$. Near the $Z$ pole, the
amplitude for $e^+e^- \to f\bar{f}$ can be written as a Laurent expansion about
the complex pole $s_0 \equiv \mz^2 - i\mz \gz$,
\begin{equation}
{\cal A}[e^+e^- \to f \bar{f}] = \frac{R}{s-s_0} + S +
        (s-s_0) S' +
\dots, \label{polexp}
\end{equation}
where $\mz$ and $\gz$ are the on-shell mass and width of the $Z$ boson,
respectively. It has been shown \cite{prop} that the coefficients $R,S,S',\dots$
and the pole location $s_0$ are individually gauge-invariant, UV- and IR-finite,
when soft and collinear real photon and gluon emission is included.

Note that, based on eq.~\eqref{polexp}, the $s$-dependence of the
cross-section near the $Z$ pole is
given by $\sigma \propto [(s-\mz^2)^2 + \mz^2\gz^2]^{-1}$, whereas in
experimental analyses a Breit-Wigner function with a running (energy-dependent)
width is being employed, $\sigma \propto [(s-\MZ^2)^2 + s^2\GZ^2/\MZ^2]^{-1}$.
Due to these different parametrizations, the experimental mass and width, $\MZ$
and $\GZ$, differ from the pole mass and width, $\mz$ and $\gz$, by a fixed
factor:
\begin{equation}
\textstyle
\mz = \MZ\big/\sqrt{1+\Gamma_\PZ^2/\MZ^2}\,, \qquad
\gz = \Gamma_\PZ\big/\sqrt{1+\Gamma_\PZ^2/\MZ^2}\,. \label{massrel}
\end{equation}
Numerically, this leads to $\mz \approx \MZ - 34\mev$ and $\gz \approx
\Gamma_\PZ - 0.9\mev$.

\vspace{\bigskipamount}
The total width, $\gz$, is related to the imaginary part of the complex pole
$s_0$. It can be obtained by requiring that the $Z$ propagator has a pole for $s=s_0$, $i.\,e.$
\begin{equation}
s_0-\mz^2+\Sigma_\PZ(s_0)=0,
\end{equation}
where $\Sigma_\PZ(s)$ is the transverse part of the $Z$ self-energy. The real
and imaginary part of this equation, respectively, lead to
\refstepcounter{equation}\label{gz}
\begin{equation}
\text{Re}\,\Sigma_\PZ(s_0)=0, \qquad\qquad
\gz = \frac{1}{\mz} \text{Im}\,\Sigma_\PZ(s_0).
\tag{\theequation a,b}
\end{equation}
Expanding eq.~(\ref{gz}b) up to next-to-next-to-leading order in $\alpha$, with
the power counting $\gz \sim \OO(\alpha)$, and using eq.~(\ref{gz}a), one obtains
\begin{align}
\gz = \frac{1}{\mz}\,
\Bigl \{&\text{Im}\,\Sigma_{\PZ(1)} 
\;+\; \text{Im}\,\Sigma_{\PZ(2)} - 
 (\text{Im}\,\Sigma_{\PZ(1)})(\text{Re}\,\Sigma'_{\PZ(1)}) \nonumber\\
&+ \text{Im}\,\Sigma_{\PZ(3)} + (\text{Im}\,\Sigma_{\PZ(1)})
  \bigl [(\text{Re}\,\Sigma'_{\PZ(1)})^2-\text{Re}\,\Sigma'_{\PZ(2)} \bigr ] -
 (\text{Im}\,\Sigma_{\PZ(2)})(\text{Re}\,\Sigma'_{\PZ(1)}) \\
&-\tfrac{1}{2}\mz \gz (\text{Im}\,\Sigma_{\PZ(1)})(\text{Im}\,\Sigma''_{\PZ(1)})
\Bigr \}_{s=\mz^2}\;.  \nonumber
\end{align}
Here the subscripts in brackets indicate the loop order and
$\Sigma'_\PZ$ is the derivative of $\Sigma_\PZ$. Making use of optical theorem,
one can relate the imaginary part of the self-energy to the decay process
$Z \to f\bar{f}$, which gives
\begin{equation}
\text{Im}\,\Sigma_{\PZ} = \frac{1}{3\mz} \sum_{f} \sum_{\rm
spins} \int d\Phi \;\bigl (|v_f|^2 + |a_f|^2\bigr ),
\end{equation}
where $v_f$ and $a_f$ are the effective vector and axial-vector couplings,
respectively, of the $Zf\bar{f}$ vertex, which include EW vertex corrections
and  $Z$--$\gamma$ mixing contributions. Final-state QED and QCD corrections can
be added via factorized radiator functions ${\cal R}_{\rm V,A}$, so that
one arrives at
\begin{align}
\gz &= \sum_f \overline{\Gamma}_f\,, \qquad
\overline{\Gamma}_f = \frac{N_c^f\mz}{12\pi} \Bigl [
 {\cal R}_{\rm V}^f F_{\rm V}^f + {\cal R}_{\rm A}^f F_{\rm A}^f \Bigr ]_{s=\mz^2} 
 \;, \label{Gz} \\
F_{\rm V}^f &= v_{f(0)}^2
 \bigl [1-\text{Re}\,\Sigma'_{\PZ(1)}-\text{Re}\,\Sigma'_{\PZ(2)}
  + (\text{Re}\,\Sigma'_{\PZ(1)})^2 \bigr ] 
 + 2 \,\text{Re}\, (v_{f(0)}v_{f(1)})\bigl [1-\text{Re}\,\Sigma'_{\PZ(1)} \bigr ] 
\nonumber \\[.5ex] 
 &\quad + 2 \,\text{Re}\, (v_{f(0)}v_{f(2)}) + |v_{f(1)}|^2
 - \tfrac{1}{2}\mz \gz v_{f(0)}^2
 \;\text{Im}\,\Sigma''_{\PZ(1)}\;, \label{Fv} \\[1ex]
F_{\rm A}^f &= a_{f(0)}^2
 \bigl [1-\text{Re}\,\Sigma'_{\PZ(1)}-\text{Re}\,\Sigma'_{\PZ(2)}
  + (\text{Re}\,\Sigma'_{\PZ(1)})^2 \bigr ] 
 + 2 \,\text{Re}\, (a_{f(0)}a_{f(1)})\bigl [1-\text{Re}\,\Sigma'_{\PZ(1)} \bigr ] 
\nonumber \\[.5ex] 
 &\quad + 2 \,\text{Re}\, (a_{f(0)}a_{f(2)}) + |a_{f(1)}|^2
 - \tfrac{1}{2}\mz \gz a_{f(0)}^2
 \;\text{Im}\,\Sigma''_{\PZ(1)}\;, \label{Fa}
\end{align}
where $N_c^f = 3(1)$ for quarks (leptons).
The functions ${\cal R}_{\rm V,A}$ have been computed including higher-order QCD
corrections up to ${\cal O}(\as^4)$ in the limit of massless quarks and ${\cal
O}(\as^3)$ for the kinematic mass corrections \cite{rad,rad2}. Furthermore,
$\OO(\alpha^2)$ QED corrections have been obtained in Ref.~\cite{Kataev:1992dg}.
The complete expressions used in this work are given in the appendix.

However, the factorization between final-state QCD/QED corrections and
EW loop corrections from massive gauge-boson exchange is not exact, but
there are additional non-factorizable contributions from irreducible vertex
diagrams. The leading non-factor\-izable corrections at ${\cal O}(\alpha\as)$
have been computed in Ref.~\cite{nfact,nfactb}, and were found to be relatively small,
but not negligible compared to the current experimental uncertainty.

%%%%%%%%%%%%%%%%%%%%%%%%%%%%%%%%%%%%%%%%%%%%%%%%%%%%%%%%%%%%%%
\begin{figure}[tb]
\begin{tabular}{@{}ccc@{}}
\psfig{figure=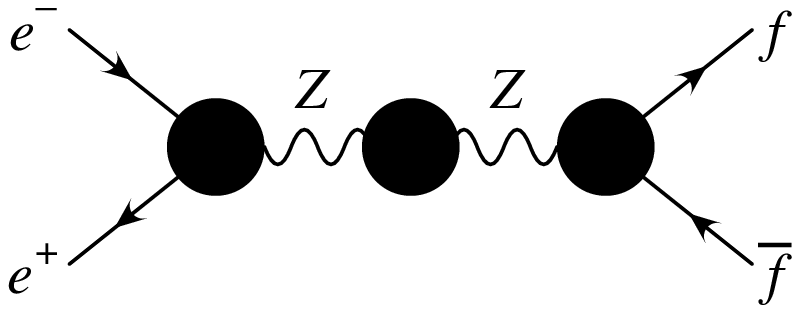, width=4.6cm} &
\psfig{figure=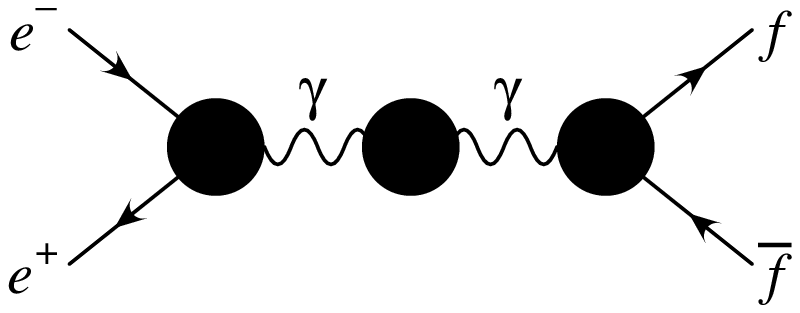, width=4.6cm} &
\raisebox{-4mm}{\psfig{figure=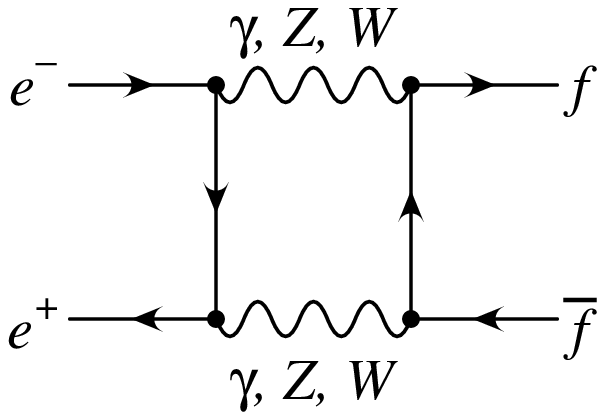, width=3.4cm}}%
\raisebox{-1mm}{\psfig{figure=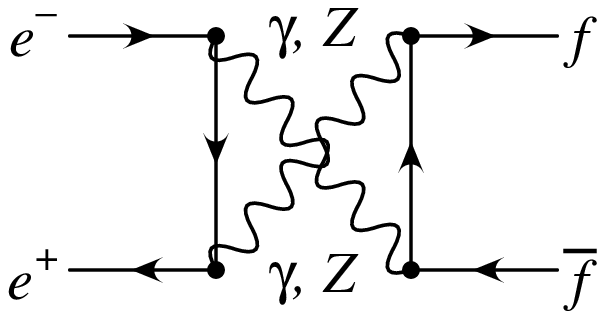, width=3.4cm}} \\[-1.5ex]
 (a) & (b) & (c)
\end{tabular}
\mycaption{Generic diagrams for $s$-channel $Z$-boson (a) and photon (b) exchange,
as well as box graphs (c). The blobs indicate self-energy and vertex loop
corrections.
\label{fig:diag1}}
\end{figure}
%%%%%%%%%%%%%%%%%%%%%%%%%%%%%%%%%%%%%%%%%%%%%%%%%%%%%%%%%%%%%%

\vspace{\bigskipamount}
The hadronic peak cross section, $\sigma^0_{\rm had}$, is phenomenologically
defined as the total cross section for $e^+e^- \to (Z) \to \text{hadrons}$ for
$s=\MZ^2$, after subtraction of $s$-channel photon exchange and box diagram 
contributions, see Fig.~\ref{fig:diag1}~(b,c), and
de-convolution of initial-state QED radiation \cite{zpole,lep1}. The notation
$(Z)$ in brackets is supposed to indicate that the unstable $Z$-boson is not an
asymptotic state. As mentioned above,
the experimental collaborations commonly parametrize the cross section,
$\sigma_{\rm had} = \sigma[e^+e^- \to (Z) \to \text{hadrons}]$ near the $Z$
peak through a Breit-Wigner form with an energy-dependent width,
\begin{equation}
\sigma_{\rm had}(s) = \sigma^0_{\rm had} \frac{s\GZ^2}{(s-\MZ^2)^2+s^2\GZ^2/\MZ^2},
\label{sh0}
\end{equation}
so that $\sigma^0_{\rm had}=\sigma_{\rm had}(\MZ^2)$. Note that one obtains the
same result at the location of the complex pole mass: $\sigma^0_{\rm
had}=\sigma_{\rm had}(\mz^2)$.

On the theory side, the hadronic peak cross section is computed from the amplitude in
eq.~\eqref{polexp}. The latter can be written as ${\cal A}(s) = {\cal A}_\PZ(s) + {\cal
A}_\gamma(s) + B(s)$, where ${\cal A}_\PZ$ and ${\cal A}_\gamma$ are the
terms from $Z$-boson and photon exchange, respectively, and $B$ denotes
EW box diagram contributions. Then
\begin{equation}
\sigma_{\rm had} = \frac{1}{64\pi^2s}
 \sum_{f=u,d,c,s,b}\int d\Omega\;\bigl|{\cal A}_\PZ(s) 
 \bigr|^2 .
\end{equation}
% prefactor checked
Expanding ${\cal A}_\PZ$ about the complex pole $s_0$ as in
eq.~\eqref{polexp}, the gauge-invariant coefficients read, including
electroweak next-to-next-to-leading order corrections (final-state radiation
will be added later),
\begin{align}
R_\PZ &= \begin{aligned}[t]
&z^\mu_{e(0)} z^\mu_{f(0)} \bigl [1-\Sigma'_{\PZ(1)}
 - \Sigma'_{\PZ(2)} + (\Sigma'_{\PZ(1)})^2 + i\mz\gz \Sigma''_{\PZ(1)}\bigr ] \\
&+\bigl [z^\mu_{e(1)} z^\mu_{f(0)} + z^\mu_{e(0)} z^\mu_{f(1)} \bigr ]
\bigl [1-\Sigma'_{\PZ(1)} \bigr ] 
+ z^\mu_{e(2)} z^\mu_{f(0)} + z^\mu_{e(0)} z^\mu_{f(2)} 
+ z^\mu_{e(1)}z^\mu_{f(1)}  \\
&-i\mz\gz \bigl [z^{\mu\,\prime}_{e(1)} z^\mu_{f(0)} + z^\mu_{e(0)}
 z^{\mu\,\prime}_{f(1)} \bigr ],
\end{aligned} \label{re} \displaybreak[0] \\[1ex]
S_\PZ &= z^{\mu\,\prime}_{e(1)} z^\mu_{f(0)} + z^\mu_{e(0)} z^{\mu\,\prime}_{f(1)}
 + \tfrac{1}{2}z^\mu_{e(0)} z^\mu_{f(0)} \Sigma''_{\PZ(1)}, \label{ae}  \\[1ex]
S'_\PZ &= 0. \label{aep}
\end{align}
Here the consistent power counting $\gz = {\cal O}(\alpha)$ has been used, so
that $R_\PZ$ is needed to two-loop order, while it is sufficient to compute $S$
and $S'$ to one-loop and tree-level order, respectively. $z^\mu_{f(n)}$ denotes
the $n$-loop correction to the effective $Zf\bar{f}$ vertex, $i.\,e.$
\begin{equation}
z^\mu_{f(n)} = v_{f(n)}\gamma^\mu + a_{f(n)}\gamma^\mu\gamma^5.
\end{equation}
Inserting the expressions in eqs.~\eqref{re}--\eqref{aep} into
eq.~\eqref{sh0} and setting $s=\mz^2$, one finds
\begin{equation}
\sigma^0_{\rm had} = \sum_{f=u,d,c,s,b} \frac{N_c^f}{12\pi\gz^2}
 \Bigl [\bigl (F_{\rm V}^e + F_{\rm A}^e\bigr )
  \bigl (F_{\rm V}^f + F_{\rm A}^f\bigr ) + 
  \bigl (v_{e(0)}^2 + a_{e(0)}^2\bigr )
  \bigl (v_{f(0)}^2 + a_{f(0)}^2\bigr )\delta X_{(2)}
  \Bigr]_{s=\mz^2}\,,
\end{equation} 
where
\begin{equation}
\delta X_{(2)} = -(\text{Im}\,\Sigma'_{\PZ(1)})^2 - 2\gz\mz \;\text{Im}\,\Sigma''_{\PZ(1)}
\label{dx}\,.
\end{equation} 
So far, this expression for $\sigma^0_{\rm had}$ only accounts for the
virtual EW corrections. As for the total width, final-state
QED and QCD radiation can be included through the radiator functions ${\cal
R}_{\rm V,A}$. The final result for $\sigma^0_{\rm had}$ can then be written as
\begin{equation}
\sigma^0_{\rm had} = \sum_{f=u,d,c,s,b}
 \frac{12\pi}{\mz^2}\,\frac{\overline{\Gamma}_e\overline{\Gamma}_f}{\gz^2}
 (1+\delta X)\,.
\end{equation} 
The correction factor $\delta X$ occurs first at two-loop level, see
eq.~\eqref{dx}, and can be traced to the last two terms in the first line of
eq.~\eqref{re}. Its existence has been realized earlier in
Ref.~\cite{Grassi:2000dz}, although eq.~\eqref{dx} differs from the expression
given there. This difference stems from the non-resonant term eq.~\eqref{ae},
which has not been included in Ref.~\cite{Grassi:2000dz}.

%%%%%%%%%%%%%%%%%%%%%%%%%%%%%%%%%%%%%%%%%%%%%%%%%%%%%%%%%%%%%%
\begin{figure}[tb]
\centering
\begin{tabular}{ll}
\raisebox{-1.1cm}{\psfig{figure=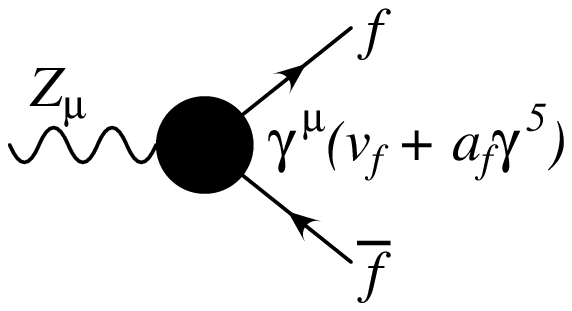, height=2.2cm}} = &
 \raisebox{-0.75cm}{\psfig{figure=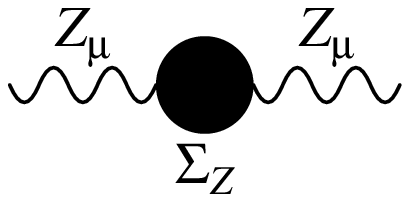, height=1.5cm}} = \\
\phantom{+} \psfig{figure=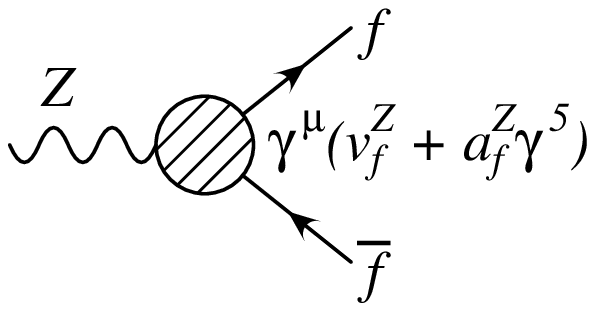, height=2.2cm} &
 \phantom{+} \raisebox{0.35cm}{\psfig{figure=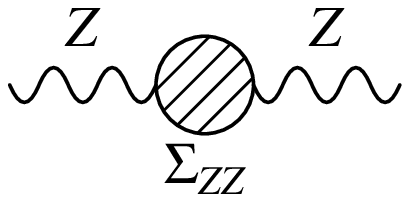, height=1.5cm}} \\
+ \raisebox{-1.1cm}{\psfig{figure=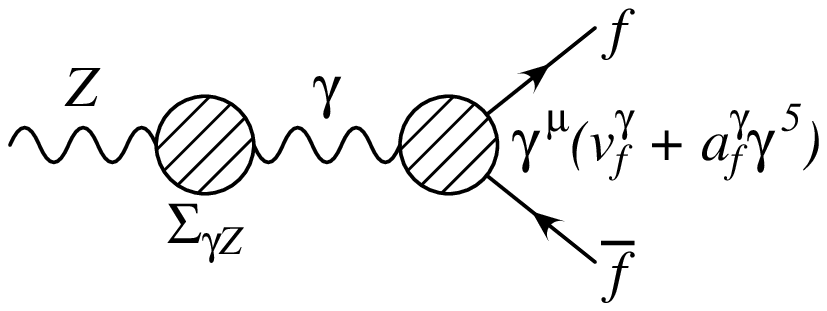, height=2.2cm}} &
 + \raisebox{-0.75cm}{\psfig{figure=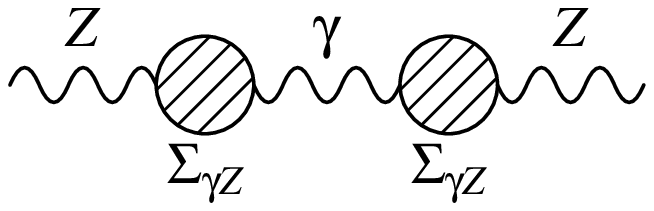, height=1.5cm}} \\ 
+ \raisebox{-1.1cm}{\psfig{figure=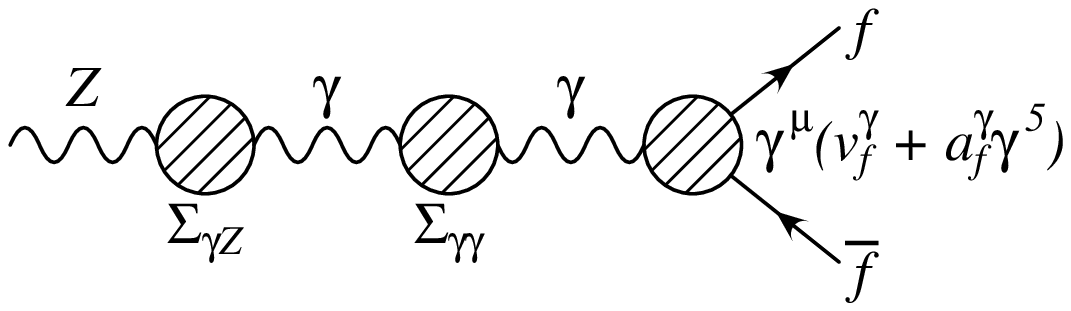, height=2.2cm}}\hspace{-2cm} &
 + \raisebox{-0.75cm}{\psfig{figure=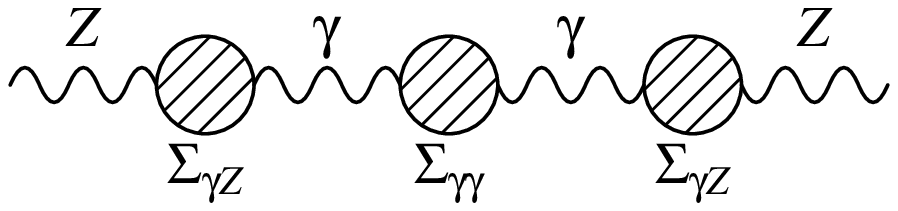, height=1.5cm}} \\ 
+ \dots & + \dots
\end{tabular}
\mycaption{Decomposition of the effective $Zf\bar{f}$ vertex and $Z$ self-energy
into one-particle irreducible building blocks, indicated by the
hatched blobs.
\label{fig:diag2}}
\end{figure}
%%%%%%%%%%%%%%%%%%%%%%%%%%%%%%%%%%%%%%%%%%%%%%%%%%%%%%%%%%%%%%

Note that $v_f$, $a_f$ and $\Sigma_Z$ as defined above include $\gamma$--$Z$
mixing contributions, see Fig.~\ref{fig:diag2}. Specifically,
\begin{align}
v_f(s) &= v_f^\PZ(s) - v_f^\gamma(s)\, 
 \frac{\Sigma_{\gamma Z}(s)}{s+\Sigma_{\gamma\gamma}(s)}, \\
a_f(s) &= a_f^\PZ(s) - a_f^\gamma(s)\, 
 \frac{\Sigma_{\gamma Z}(s)}{s+\Sigma_{\gamma\gamma}(s)}, \\
\Sigma_Z(s) &= \Sigma_{ZZ}(s) - \frac{[\Sigma_{\gamma
Z}(s)]^2}{s+\Sigma_{\gamma\gamma}(s)},
\end{align}
where $v_f^\PZ$ and $a_f^\PZ$ are the one-particle irreducible $Zf\bar{f}$
vector and axial-vector vertex contributions, respectively, and $v_f^\gamma$
and $a_f^\gamma$ are their equivalent for the $\gamma f\bar{f}$ vertex.
The symbols $\Sigma_{V_1V_2}$ denote the one-particle irreducible corrections to
the $V_1$--$V_2$ self-energy.

%%%%%%%%%%%%%%%%%%%%%%%%%%%%%%%%%%%%%%%%%%%%%%%%%%%%%%%%%%%%%%

\section{Calculation of fermionic two-loop corrections}
\label{sc:calc}

In the calculation of the electroweak two-loop corrections with one or two
closed fermion loops, the masses and Yukawa couplings of all fermions except the
top quark have been neglected. [For the one-loop and $\OO(\alpha\as)$
corrections, the finite bottom-quark mass has been retained.] Moreover, the
quark mixing matrix is assumed to be diagonal. The diagrams for the $Zf\bar{f}$
vertex corrections and for the renormalization terms have been generated with
{\sc  FeynArts 3.3} \cite{feynarts}. Dimensional regularization is used for defining
potentially UV-divergent loop integrals.
The vector and axial-vector form factors,
$v_{f(n)}$ and $a_{f(n)}$, have been singled out by contraction with suitable
projection operators,
\begin{align}
v_\Pf(k^2) &= \frac{1}{2(2-d)k^2} \, {\rm Tr}[\gamma_\mu \, \pslash_1 \,
z_f^\mu(k^2) \, \pslash_2], \\
a_\Pf(k^2) &= \frac{1}{2(2-d)k^2} \, {\rm Tr}[\gamma_5 \,
\gamma_\mu \, \pslash_1 \,
z_f^\mu(k^2) \, \pslash_2],
\end{align}
where $d$ is the space-time dimension and $p_{1,2}$ are the momenta of the
external fermions. The resulting expression contain only scalar loop integrals,
which however may still contain non-trivial structures of scalar products in their
numerators.

For the purpose of the work presented here, the on-shell renormalization scheme
is employed. In this scheme, the renormalized squared masses are defined at the
real part of the propagator poles, see also eq.~\eqref{polexp}. Specifically,
for the $\OO(\alpha^2)$ corrections, mass counterterms are needed for $\mw$,
$\mz$, and $\mt$. The electromagnetic charge is defined as the coupling
strength of the $\gamma f\bar{f}$ in the limit of zero photon momentum, and the
\emph{on-shell} weak mixing angle is defined through the ratio of the
renormalized $W$ and $Z$ masses, $\sw = 1-\mw^2/\mz^2$. Finally, wave function
renormalization constants for the external fermions are needed. Detailed
expressions for the relevant counterterms are given in Ref.~\cite{mwlong}. When
computing the $Z$ decay width, one cannot define a physical
wave-function or field-strength renormalization of the incoming $Z$ boson, since
it is unstable and thus not an asymptotic state\footnote{For a technical
definition of the of the field-strength renormalization of unstable particles,
see $e.\,g.$ Ref.~\cite{nekrasov}.}. Instead, UV-divergencies associated in the
incoming $Z$-boson line are canceled by the terms involving the derivative of
the self-energy, $\Sigma'_\PZ$, in eqs.~\eqref{Fv}, \eqref{Fa}.

\vspace{\bigskipamount}
Two-loop self-energy integrals and vertex integrals with sub-loop self-energy
bubbles have been evaluated with the method illustrated in section~3.2 of
Ref.~\cite{swlept2}. In this approach, the loop integrals are reduced to a small
set of master integrals using a generalization of Passarino-Veltman reduction
\cite{weiglein}, as well as integration-by-parts \cite{ibp} and Lorentz identities
\cite{li}. The master integrals are then evaluated using very efficient one-dimensional
numerical integrations \cite{bauberger, swlept2}.

For the computation of $\Sigma'_\PZ$, derivatives of self-energy integrals are
needed. To illustrate their evaluation, let us write a two-loop self-energy master integral
in the form
\begin{multline}
T(p^2;m_1^2,m_2^2,m_3^2,m_4^2,m_5^2;\nu_1,\nu_2,\nu_3,\nu_4,\nu_5) 
 = -\frac{(4\pi^2\mu^2)^{4-d}}{\pi^4} \\ \times\int \frac{d^dq_1 d^dq_2}{[q_1^2-m_1^2]^{\nu_1}
 [(q_1+p)^2-m_2^2]^{\nu_2}[(q_1-q_2)^2-m_3^2]^{\nu_3}
 [q_2^2-m_4^2]^{\nu_4}[(q_2+p)^2-m_5^2]^{\nu_5}} %\nonumber
\end{multline}
where $p$ is the external momentum. The derivative with respect to $p^2$ can be
expressed in terms of the same functions:
\begin{multline}
\frac{\partial T}{\partial(p^2)} = \frac{1}{2p^2} p^\mu \frac{\partial T}{\partial
p^\mu} 
 = -\frac{1}{2p^2}\bigl [(\nu_2+\nu_5)T -\nu_2T(\nu_1-1,\nu_2+1) 
   - \nu_5T(\nu_4-1,\nu_5+1) \\
 + \nu_2(m_2^2-m_1^2+p^2)T(\nu_2+1)
 + \nu_5(m_5^2-m_4^2+p^2)T(\nu_5+1)\bigr ].
\end{multline}
With the help of integration-by-parts identities, the resulting integrals can
again be reduced to a basic set with all $\nu_i$ either 1 or 0.

%%%%%%%%%%%%%%%%%%%%%%%%%%%%%%%%%%%%%%%%%%%%%%%%%%%%%%%%%%%%%%
\begin{figure}[tb]
\centering
\psfig{figure=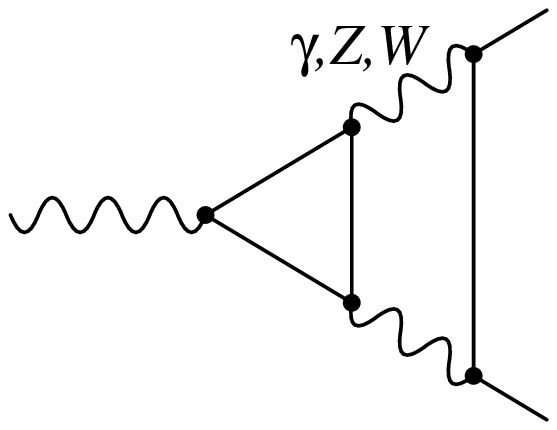, width=3.9cm} 
\hspace{1cm}
\raisebox{1em}{\begin{minipage}{8cm}
\mycaption{Generic two-loop vertex diagram with triangle fermion sub-loop.}
\end{minipage}}
\label{fig:tri}
\end{figure}
%%%%%%%%%%%%%%%%%%%%%%%%%%%%%%%%%%%%%%%%%%%%%%%%%%%%%%%%%%%%%%

\vspace{\bigskipamount}
For vertex diagrams with sub-loop triangles, see Fig.~\ref{fig:tri}, the numerical
integration technique of Ref.~\cite{sub} has been used. This method is based on a
direct integration in Feynman parameter space, without prior tensor reduction,
and using a deformation of integration contours into the complex plane to avoid
poles from physical thresholds. Where applicable, results for individual
diagrams obtained with this approach have been compared to Ref.~\cite{swbb}.

In this context, it should be pointed out that diagrams with fermion triangle
sub-loop involve chiral couplings that lead to Dirac traces of the form
\begin{equation}
\text{tr}\{\gamma^\alpha\gamma^\beta\gamma^\gamma\gamma^\delta\gamma_5\} = 4i
\epsilon^{\alpha\beta\gamma\delta}. \label{trg5}
\end{equation}
Within dimensional regularization, eq.~\eqref{trg5} is inconsistent with the
anticommutation rule $\{\gamma^\mu,\gamma_5\}=0$. However, it can be shown (see
for instance Ref.~\cite{jegg5}) that
contributions proportional to epsilon tensors of the form \eqref{trg5} are
separately gauge-invariant and UV-finite, and thus they can be computed in four
dimensions with a well-defined result. In order to avoid difficulties due to
spurious IR-singularities from diagrams with massless photons, a small photon
mass is introduced \cite{swlept2}. The sum of all triangle
sub-loop  diagrams is IR-finite, so that any remaining contribution from the
small photon mass is power suppressed.

\vspace{\bigskipamount}
Additional checks have been performed for the complete result, combining all
building blocks, including two-loop vertex diagrams, counterterms for the
renormalization and reducible terms ($i.\,e.$ $\OO(\alpha^2)$ contributions that
factorize into a product of one-loop factors). The UV- and IR-finiteness has
been checked analytically within dimensional regularization. The final
expressions
for the vector and axial-vector form factors, $v_{f(2)}$ and $a_{f(2)}$, can
also be used to evaluate the effective weak mixing angle $\sin^2 \theta_{\rm
eff}^f = \frac{1}{4}(1+\text{Re}\{v_l/a_l\})$, and very good agreement with the
literature has been obtained for $f=\ell$ \cite{swlept} and $f=b$
\cite{swbb}. Furthermore, the fermionic two-loop result for $R_b$ from
Ref.~\cite{rb} has been reproduced with good accuracy. In all of these
comparisons, the agreement is better than the intrinsic uncertainty from
numerical integration errors.

%Note that in these
%comparisons the numerical agreement is not perfect, since in Refs.~\cite{swlept,swbb,rb}
%perturbation theory was applied directly to the given observable, whereas in
%this work the form factors $v_f$ and $a_f$ were evaluated separately and then
%combined numerically. The two approaches differ by higher-order terms, so that
%the discrepancy can be identified as part of the theoretical uncertainty, see
%also section~\ref{sc:error}.

%%%%%%%%%%%%%%%%%%%%%%%%%%%%%%%%%%%%%%%%%%%%%%%%%%%%%%%%%%%%%%

\section{Numerical results}
\label{sc:res}

In this section, numerical results for the partial widths $\Gamma_f \equiv
\Gamma(Z \to f\bar{f})$, their ratios, the total $Z$ width $\GZ$, and the
hadronic peak cross-section $\sigma_{\rm had}$ are presented. Although the
calculation is performed using the complex-pole definition of the gauge-boson
masses, see section~\ref{sc:def}, the numerical results are given in terms of
the running width scheme, $i.\,e.$ \MZ\ and \GZ\ in eq.~\eqref{massrel}.

All currently known perturbative corrections are included in the result:
\begin{itemize}
\item $\OO(\alpha)$ and and fermionic $\OO(\alpha^2)$ EW contributions (from this
work); 
\item $\OO(\alpha\as)$ corrections to internal gauge-boson self-energies
\cite{qcd2} (which have been re-computed for this work);
\item leading three- and four-loop corrections in the large-$\mt$ limit, of
order $\OO(\alpha_\Pt\as^2)$ \cite{qcd3}, $\OO(\alpha_\Pt^2\as)$, $\OO(\alpha_\Pt^3)$ \cite{mt6}, and $\OO(\alpha_\Pt\as^3)$ 
\cite{qcd4}, where $\alpha_\Pt \equiv \alpha \mt^2$;
\item final-state QED and QCD (for quark final states) radiation up to
$\OO(\alpha^2)$, $\OO(\alpha\as)$ and $\OO(\as^4)$ \cite{rad,rad2,Kataev:1992dg},
which are incorporated through the radiator functions ${\cal R}_{\rm V,A}$;
\item non-factorizable $\OO(\alpha\as)$ vertex contributions \cite{nfact,nfactb}, which
account for the fact that the factorization between 
EW corrections in $F_{\rm V,A}$ and final-state radiation effects in ${\cal
R}_{\rm V,A}$ is not exact.
\end{itemize}
Light fermion masses $\mf$, $f\neq t$, have been neglected everywhere except for
a non-zero $b$ quark mass in the $\OO(\alpha)$ and
$\OO(\alpha\as)$ contributions, non-zero bottom, charm and tau masses in the
radiators ${\cal R}_{\rm V,A}$.  The top-quark mass, $\mt$, has been defined in the
on-shell scheme, while the $\overline{\text{MS}}$-scheme has been used for
$m_{\rm c}$ and $\mb$.

%-------------------------------------------------------------
\begin{table}[tb]
\renewcommand{\arraystretch}{1.2}
\begin{center}
\begin{tabular}{|ll|ll|}
\hline
Parameter & Value & Parameter & Value \\
\hline \hline
$\MZ$ & 91.1876 GeV & $\mb^{\overline{\rm MS}}$ & 4.20 GeV \\
$\Gamma_\PZ$ & 2.4952 GeV & $m_{\rm c}^{\overline{\rm MS}}$ & 1.275 GeV \\
$\MW$ & 80.385 GeV & $m_\tau$ & 1.777 GeV \\
$\Gamma_\PW$ & 2.085 GeV & $\Delta\alpha$ & 0.05900 \\
$\MH$ & 125.7 GeV & $\as(\MZ)$ & 0.1184 \\
$\mt$ & 173.2 GeV & $G_\mu$ & $1.16638 \times 10^{-5}$~GeV$^{-2}$ \\
\hline
\end{tabular}
\end{center}
\vspace{-2ex}
\mycaption{Input parameters used in the numerical analysis,
from Refs.~\cite{pdg,gfit,dalept,dahad}. 
%$\Delta\alpha$ accounts for the running of the 
%electromagnetic coupling due to loop corrections from leptons
%\cite{dalept} and the five light quark flavors, $\Delta\alpha=\Delta\alpha_{\rm
%lept}(\MZ)+\Delta\alpha^{(5)}_{\rm had}(\MZ)$.
\label{tab:input}}
\end{table}
%-------------------------------------------------------------

Owing to the renormalization scheme used here, the EW corrections are
obtained in terms of the electromagnetic coupling $\alpha$, rather than the
Fermi constant $G_\mu$, as expansion parameter. However, in a second step,  the
measured value of $G_\mu$ is used to compute $\MW$ within the SM, leading to a
prediction of the (partial) $Z$ width and cross-section in terms of $\MZ$,
$\MH$, $\mt$, $\mb^{\overline{\rm MS}}$, $m_{\rm c}^{\overline{\rm MS}}$,
$m_\tau$, $G_\mu$, $\alpha$, $\as$ and $\Delta\alpha$. Here $\Delta\alpha$
describes the shift in the electromagnetic coupling between the scales $q^2=0$
and $\MZ^2$ due to light fermion loops, $\alpha(\MZ^2) =
\alpha(0)/(1-\Delta\alpha)$. While the contribution to $\Delta\alpha$
from leptons has been computed perturbatively up to three-loop level
\cite{dalept}, $\Delta\alpha_{\rm
lept}(\MZ) = 0.0314976$, the quark loops at low scales lead to non-perturbative
contributions that have to be taken from experimental data, see $e.\,g.$
\cite{dahad}, and the value $\Delta\alpha^{(5)}_{\rm had}(\MZ) = 0.02750$ is
adopted here. The numerical input values used in this section are listed in
Tab.~\ref{tab:input}.

\vspace{\bigskipamount}
Results for the total width, $\GZ$, and hadronic peak cross-section,
$\sigma^0_{\rm had}$, have been presented in Ref.~\cite{short}. In the following
subsections, numerical results for the partial widths and branching ratios will
be discussed.

\subsection{Partial widths}

Results for the contribution of the different loop orders to various partial widths
are shown in Tab.~\ref{tab:res1} for a fixed value of $\MW$ ($i.\,e.$ $G_\mu$
is not used as an input parameter here). As evident from the table, the two-loop
EW corrections are sizeable, of the same order as the $\OO(\alpha\as)$ terms.

%-------------------------------------------------------------
\begin{table}[tbp]
\renewcommand{\arraystretch}{1.2}
\begin{center}
\begin{tabular}{|l|r|r|r|r|r|r|}
\hline
\multicolumn{1}{|r|}{$\Gamma_i$ [MeV]} & $\Gamma_e\;\;$ & $\Gamma_\nu\;\;$ & $\Gamma_d\;\;$ & $\Gamma_u\;\;$ & 
 $\Gamma_b\;\;$ & $\Gamma_\PZ\;\;$ \\
\hline \hline
$\OO(\alpha)$ & 2.274 & 6.176 & 9.724 & 5.804 & 3.863 & 60.26 \\
$\OO(\alpha\as)$ & 0.288 & 0.458 & 1.276 & 1.156 & 2.006 & 9.11 \\
$\OO(\alpha_\Pt\as^2,\,\alpha_\Pt\as^3,\,\alpha^2_\Pt\as,\,\alpha_\Pt^3)$ &
 0.038 & 0.059 & 0.191 & 0.170 & 0.190 & 1.20 \\
$\OO(N_f^2\alpha^2)$ & %0.228 & 0.416 & 0.693 & 0.506 & 0.689 & 5.02 & $-6.64$ \\
                      0.244 & 0.416 & 0.698 & 0.528 & 0.694 & 5.13 \\
$\OO(N_f\alpha^2)$ & %0.112 & 0.186 & 0.490 & 0.482 & 0.148 & 2.99 & 6.06 \\
                    0.121 & 0.186 & 0.494 & 0.494 & 0.144 & 3.04 \\
\hline
\end{tabular}
\end{center}
\vspace{-2ex}
\mycaption{Loop contributions, in units of MeV, to the partial and total $Z$ widths with
fixed $\MW$ as input parameter. Here $N_f$ and $N_f^2$ refer to corrections with 
one and two closed fermion loops, respectively, and $\alpha_\Pt = \alpha\mt^2$.
In all rows the radiator functions ${\cal R}_{\rm V,A}$ with known contributions
through $\OO(\as^4)$, $\OO(\alpha^2)$ and $\OO(\alpha\as)$ are included.
\label{tab:res1}}
\end{table}
%-------------------------------------------------------------

\begin{figure}[tbp]
\begin{minipage}[t]{8cm}
\epsfig{figure=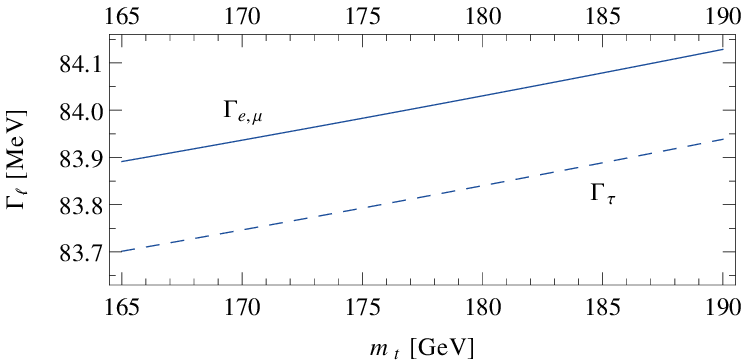, width=8cm, bb=0 36 360 177, clip=true}\newline
\anc\hfill\epsfig{figure=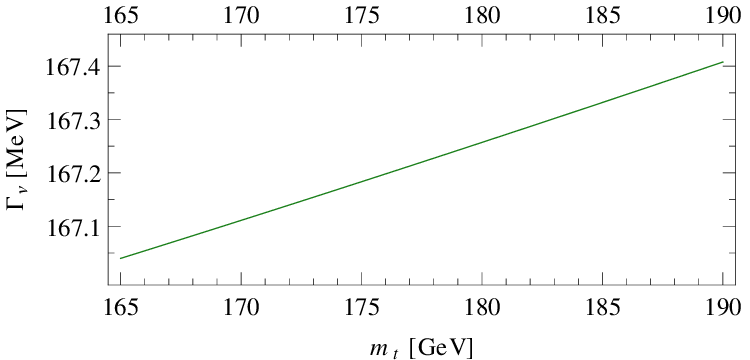, width=7.98cm, bb=0 36 360 157, clip=true}\newline
\epsfig{figure=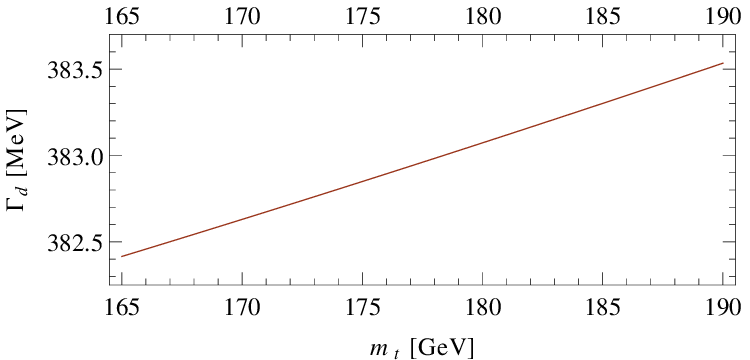, width=8cm, bb=0 0 360 157, clip=true}%
\end{minipage}
\begin{minipage}[t]{8cm}
\anc\hfill
\epsfig{figure=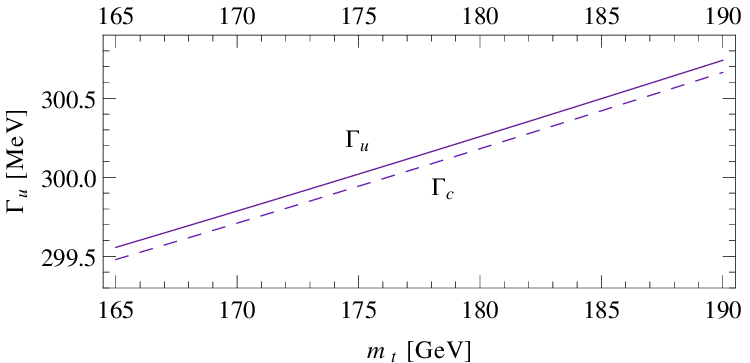, width=8cm, bb=0 36 360 174, clip=true}\newline
\epsfig{figure=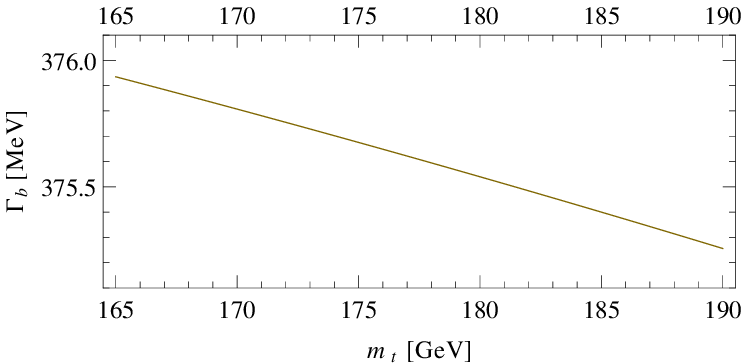, width=8cm, bb=0 0 360 158, clip=true}
\end{minipage}
\mycaption{Results for the $Z$ partial widths $\Gamma_i$, with $\MW$ calculated
from $G_\mu$, using all perturbative corrections discussed in the text and
including the full radiator functions ${\cal R}_{\rm V,A}$. The dependence on
$\mt$ is shown explicitly, while the other input parameters are fixed to the
values in Tab.~\ref{tab:input}.
\label{fig:res1}}
\end{figure}

If one uses $G_\mu$ as an input to compute $\MW$, using the results from
Ref.~\cite{mw,mwlong,mwtot} (which have been augmented to include the four-loop
$\OO(\alpha_\Pt\as^3)$ corrections \cite{qcd4} that became available later), the values
shown in Fig.~\ref{fig:res1} are obtained. The dependence of the partial widths
on the input parameters $\mt$, $\as$ and $\MH$ is relatively mild, leading to
variations at the per-mille level within the phenomenologically
relevant ranges $165\dots190 \gev$, $0.113\dots0.123$ and $100\dots600\gev$,
respectively.

%%%%%%%%%%%%%%%%%%%%%%%%%%%%%%%%%%%%%%%%%%%%%%%%%%%%%%%%%%%%%%
%%%%%%%%%%%%%%%%%%%%%%%%%%%%%%%%%%%%%%%%%%%%%%%%%%%%%%%%%%%%%%

To illustrate the impact of the newly calculated fermionic two-loop corrections,
Tab.~\ref{tab:rese} shows a comparison to the previously known approximation of
the EW two-loop corrections for large values of $\mt$
\cite{ewmt4,ewmt2}\footnote{The author is grateful to S.~Mishima for supplying
these numbers based on the work in Ref.~\cite{xfit}.}. The new results lead to a
relative modifications of a $\text{few}\times 10^{-4}$, with the exact value
varying depending on the final state. For the total width, the shift is
smaller, but comparable to the current experimental measurement, $\GZ = (2.4952
\pm 0.0023)\gev$ \cite{lep1}, which has a relative uncertainty of about
$10^{-3}$.

%-------------------------------------------------------------
\begin{table}[tbp]
\renewcommand{\arraystretch}{1.2}
\begin{center}
\begin{tabular}{|l|r|r|}
\hline & & \\[-1.2em]
$\Gamma_i$ [GeV] & this work & \parbox[c]{2.6cm}{Large-$\mt$ exp.\newline for EW
2-loop} \\[1.2ex]
% from Oct 7 (b):
\hline
$\Gamma_{e,\mu}$ & 0.08397 & 0.08399 \\
$\Gamma_\tau$    & 0.08378 & 0.08380 \\
$\Gamma_\nu$     & 0.16716 & 0.16722 \\
$\Gamma_u$       & 0.29995 & 0.29996 \\
$\Gamma_c$       & 0.29987 & 0.29988 \\
$\Gamma_{d,s}$   & 0.38278 & 0.38290 \\
$\Gamma_b$       & 0.37573 & 0.37577 \\
\hline
$\Gamma_\PZ$     & 2.49430 & 2.49485 \\
\hline
\end{tabular}
\end{center}
\vspace{-2ex}
\mycaption{Comparison between the result based on the full fermionic two-loop EW
corrections and the large-$\mt$ approximation \cite{ewmt4,ewmt2,xfit}, with
$\MW$ calculated from $G_\mu$ at the same level of precision in each column.  In
both cases, the complete radiator functions ${\cal R}_{\rm V,A}$ are included.
For consistency of the comparison, the relatively small $\OO(\alpha_\Pt\as^3)$
contribution has been removed in the second column, since this part is also
missing in the last column. 
\label{tab:rese}}
\end{table}
%-------------------------------------------------------------

\subsection{Ratios}

Instead of directly determining the partial widths for the different final
states, the experiments at LEP and SLC obtained values for various branching
ratios,
since this permits a better control of systematic uncertainties. The most
relevant ratios are
\begin{align}
R_\ell &= \Gamma_{\rm had}/\Gamma_\ell, &
R_c &= \Gamma_c/\Gamma_{\rm had}, &
R_b &= \Gamma_b/\Gamma_{\rm had}, \label{rat}
\end{align}
where $\Gamma_\ell = \frac{1}{3}(\Gamma_e + \Gamma_\mu + \Gamma_\tau)$, and
$\Gamma_{\rm had}$ is the hadronic partial width, which at parton level is
equivalent to $\sum_q \Gamma_q$ ($q=u,d,c,s,b$).

%-------------------------------------------------------------
\begin{table}[tb]
\renewcommand{\arraystretch}{1.2}
\begin{center}
\begin{tabular}{|l|c|c|c|}
\hline
 & $R_\ell$ & $R_c$ & $R_b$ \\
\hline \hline
Born$\,+\,\OO(\alpha)$ & 20.8031 & 0.17230 & 0.21558 \\
$+\OO(\alpha\as)$ & 20.7963 & 0.17222 & 0.21593 \\
$+\OO(\alpha_\Pt\as^2,\,\alpha_\Pt\as^3,\,\alpha^2_\Pt\as,\,\alpha_\Pt^3)$ &
20.7943 & 0.17222 & 0.21593 \\
$+\OO(N_f^2\alpha^2,N_f\alpha^2)$ & 20.7509 & 0.17223 & 0.21580 \\
\hline
$-\OO(\alpha_\Pt\as^3)$ & 20.7512 & 0.17223 & 0.21580 \\
\hline
Large-$\mt$ exp.\ for EW 2-loop & 20.7484 & 0.17220 & 0.21579 \\
\hline
\end{tabular}
\end{center}
\vspace{-2ex}
\mycaption{Results for the ratios $R_\ell$, $R_c$ and $R_b$, with $\MW$ calculated
from $G_\mu$ to the same order as indicated in each
line. In all cases, the complete radiator functions ${\cal R}_{\rm V,A}$
are included. The last two lines compare the new result with the previous
calculation using a large-$\mt$ approximation
\cite{ewmt4,ewmt2,xfit}. For consistency of the comparison, the 
$\OO(\alpha_\Pt\as^3)$ contribution is not included in either of these last two
lines.
\label{tab:resr}}
\end{table}
%-------------------------------------------------------------

Numerical results for these ratios, with different orders of radiative
corrections included, are listed in Tab.~\ref{tab:resr}. Among the three
quantities in eq.~\eqref{rat}, $R_\ell$ is shows the most significant effect of the full fermionic EW
2-loop corrections in comparison to the
large-$\mt$ approximation, with a relative shift of $\sim 1.4 \times 10^{-4}$. For
$R_b$ and $R_c$ the impact of the new corrections mostly cancels in the
ratio. The
current experimental values are $R_\ell = 20.767 \pm 0.025$, $R_c = 0.1721 \pm
0.0030$, and $R_b = 0.21629 \pm 0.00066$ \cite{lep1}.

\vspace{\bigskipamount}
Note that the numbers for $R_b$ given here differ somewhat from Ref.~\cite{rb},
which is due to two factors: Firstly, the non-factorizable
$\OO(\alpha\as)$ contributions \cite{nfact, nfactb}, as well as higher-order
corrections from Ref.~\cite{mt6,qcd4} and $\OO(\as^4)$ final-state corrections
\cite{rad2} were not included in Ref.~\cite{rb}.
Together, these account for a shift of 2--$3\times 10^{-4}$, depending on the
input parameters.
Secondly, while in Ref.~\cite{rb} the perturbative expansion
was applied directly to the ratio $R_b$, the values in Tab.~\ref{tab:resr} have
been obtained using perturbative results for $\Gamma_b$ and $\Gamma_{\rm had}$,
as explained in the previous subsection, and dividing them numerically. The two
treatments differ by higher-order terms, and thus this part of the discrepancy
should be attributed to the theoretical uncertainty (see
section \ref{sc:error}).

It is recommended to use the parametrization formula for $R_b$ given
in this paper% for future applications
, rather than the one in
Ref.~\cite{rb}, since additional higher-order contributions are included here.

\subsection{Parametrization formulae}
\label{sc:fit1}

%-------------------------------------------------------------
\begin{table}[tb]
\renewcommand{\arraystretch}{1.3}
\begin{center}
\begin{tabular}{|l|cccccccc|c|}
\hline
Observable & $X_0$ & $c_1$ & $c_2$ & $c_3$ & $c_4$ & $c_5$ & $c_6$ & $c_7$ &
max.\ dev. \\
\hline
$\Gamma_{e,\mu}$ [MeV] &
 83.966 & $-$0.047 & 0.807& $-$0.095 & $-$0.01 & 0.25 & $-$1.1 & 285 
 & $<0.001$\\
$\Gamma_{\tau}$ [MeV] &
 83.776 & $-$0.047 & 0.806& $-$0.095 & $-$0.01 & 0.25 & $-$1.1 & 285 
 & $<0.001$\\
$\Gamma_{\nu}$ [MeV] &
 167.157 & $-$0.055 & 1.26 & $-$0.19  & $-$0.02 & 0.36 & $-$0.1 & 503 
 & $<0.001$\\
$\Gamma_{u}$ [MeV] &
 299.936 & $-$0.34 &  4.07 &    14.27 &     1.6 & 1.8  & $-$11.1& 1253
 & $<0.001$\\
$\Gamma_{c}$ [MeV] &
 299.860 & $-$0.34 &  4.07 &    14.27 &     1.6 & 1.8  & $-$11.1& 1253
 & $<0.001$\\
$\Gamma_{d,s}$ [MeV] &
 382.770 & $-$0.34  & 3.83 &    10.20 & $-$2.4  & 0.67 & $-$10.1& 1469 
 & $<0.001$\\
$\Gamma_{b}$ [MeV] &
 375.724 & $-$0.30 &$-$2.28&    10.53 & $-$2.4  & 1.2  & $-$10.0& 1458
 & $<0.001$\\
$\GZ$ [MeV] & 
 2494.24 & $-$2.0   & 19.7 &    58.60 & $-$4.0  & 8.0  & $-$55.9& 9267 
 & $<0.01$\\
\hline
$R_\ell$ [$10^{-3}$] &
 20750.9 & $-$8.1   & $-$39&   732.1  & $-$44   & 5.5  & $-$358 & 11702
& $<0.1$ \\
$R_c$ [$10^{-3}$] &
 172.23 & $-$0.029  & 1.0  &    2.3   &    1.3  & 0.38 & $-$1.2 & 37
& $<0.01$ \\
$R_b$ [$10^{-3}$] &
 215.80 &  0.031   &$-$2.98& $-$1.32 & $-$0.84 & 0.035 & 0.73   & $-$18
& $<0.01$ \\
\hline
$\sigma^0_{\rm had}$ [pb] & 
 41488.4 &    3.0   & 60.9 & $-$579.4 &   38    & 7.3  &   85   & $\!\!\!\!-$86027  
 & $<0.1$\\
\hline
\end{tabular}
\end{center}
\vspace{-2ex}
\mycaption{Coefficients for the parametrization formula \eqref{par1} for various
observables ($X$). Within the ranges $\MH = 125.7\pm 2.5\gev$, $\mt = 173.2\pm 2.0\gev$,
$\as=0.1184\pm 0.0050$, $\Delta\alpha = 0.0590 \pm 0.0005$ and $\MZ = 91.1876 \pm
0.0042 \gev$, the formula approximates the full result with maximal deviations
given in the last column.
\label{tab:fit1}}
\end{table}
%-------------------------------------------------------------

For practical purposes, the complete results for the partial widths, branching ratios,
and the peak cross-sections, including all higher-order corrections listed at
the beginning of section~\ref{sc:res} and $\MW$ calculated from $G_\mu$ to the
same precision, are most easily represented by a simple
parametrization formula. Within currently allowed experimental ranges for the
input parameters, the following form provides a very good description:
\begin{align}
&X = X_0 + c_1 L_\PH + c_2 \Delta_\Pt + c_3 \Delta_{\as} + c_4 \Delta_{\as}^2
 + c_5 \Delta_{\as}\Delta_\Pt + c_6 \Delta_\alpha + c_7 \Delta_\PZ, \label{par1} \\
&L_\PH = \log\frac{\MH}{125.7\gev}, \quad
 \Delta_\Pt = \Bigl (\frac{\mt}{173.2\gev}\Bigr )^2-1, \quad
 \Delta_{\as} = \frac{\as(\MZ)}{0.1184}-1, \nonumber \\
& \Delta_\alpha = \frac{\Delta\alpha}{0.059}-1, \quad
 \Delta_\PZ = \frac{\MZ}{91.1876\gev}-1. \nonumber
\end{align}
As before, $\MH$, $\MZ$, $\mt$ and $\Delta\alpha$ are defined in the on-shell
scheme, using the running-width scheme for $\MZ$ (to be consistent with the published
experimental values), while $\as$ is defined in the $\overline{\rm MS}$ scheme.

The coefficients for the different observables discussed in the previous
subsections are given in Tab.~\ref{tab:fit1}. With these parameters, the formula
provides a very good approximation to the full result within the ranges $\MH =
125.7\pm 2.5\gev$, $\mt = 173.2\pm 2.0\gev$, $\as=0.1184\pm 0.0050$,
$\Delta\alpha = 0.0590 \pm 0.0005$ and $\MZ = 91.1876 \pm 0.0042 \gev$, with
maximal deviations as quoted in the last column of Tab.~\ref{tab:fit1}.

Extended fit formulae, which cover a larger parameter region (in
particular larger ranges for $\MH$ and $\mt$), are given in appendix
\ref{sc:fit2}.

%%%%%%%%%%%%%%%%%%%%%%%%%%%%%%%%%%%%%%%%%%%%%%%%%%%%%%%%%%%%%%

\section{Error estimate}
\label{sc:error}

The results presented in this paper have an intrinsic theoretical uncertainty
from currently unknown higher-order contributions. The most important missing
pieces are the bosonic EW $\OO(\alpha^2_{\rm bos})$ corrections (stemming from two-loop
diagrams without closed fermion loops), and $\OO(\alpha^3)$, $\OO(\alpha^2\as)$,
$\OO(\alpha\as^2)$ and $\OO(\alpha\as^3)$ corrections beyond the leading 
$\mt^n$ terms from Ref.~\cite{qcd3,mt6,qcd4}.

The second category can be estimated by assuming that the perturbation series
follows roughly a geometric series. Thus one obtains
\begin{align}
\OO(\alpha^3)-\OO(\alpha_\Pt^3) &\sim 
 \frac{\OO(\alpha^2_{\rm ferm})-\OO(\alpha_\Pt^2)}{\OO(\alpha)}
 \OO(\alpha^2_{\rm ferm}), \displaybreak[0] \\
\OO(\alpha^2\as)-\OO(\alpha_\Pt^2\as) &\sim 
 \frac{\OO(\alpha^2_{\rm ferm})-\OO(\alpha_\Pt^2)}{\OO(\alpha)}
 \OO(\alpha\as), \displaybreak[0] \\
\OO(\alpha\as^2)-\OO(\alpha_\Pt\as^2) &\sim 
 \frac{\OO(\alpha\as)-\OO(\alpha_\Pt\as)}{\OO(\alpha)}
 \OO(\alpha\as), \displaybreak[0] \\
\OO(\alpha\as^3)-\OO(\alpha_\Pt\as^3) &\sim 
 \frac{\OO(\alpha\as)-\OO(\alpha_\Pt\as)}{\OO(\alpha)}
 \OO(\alpha\as^2),
\end{align}
where the known leading large-$\mt$ approximations have been subtracted in the
numerators, and $\alpha^2_{\rm ferm}$ indicates the
fermionic EW two-loop contribution discussed in this paper, which is currently
the only known $\OO(\alpha^2)$ piece. Using these expressions, one finds for the
total $Z$ width
\begin{equation}
\begin{aligned}
&\GZ: \; &\OO(\alpha^3)-\OO(\alpha_\Pt^3) &\sim 0.26\mev, \quad &
\OO(\alpha^2\as)-\OO(\alpha_\Pt^2\as) &\sim 0.30\mev, \\
&&\OO(\alpha\as^2)-\OO(\alpha_\Pt\as^2) &\sim 0.23\mev, &
\OO(\alpha\as^3)-\OO(\alpha_\Pt\as^3) &\sim 0.035\mev.
\end{aligned}
\label{errgz}
\end{equation}
The error from the missing bosonic $\OO(\alpha^2_{\rm bos})$ contributions can
be evaluated by taking the square of the bosonic one-loop corrections. For $\GZ$
this leads to the estimate $\OO(\alpha^2_{\rm bos}) \sim 0.1\mev$. 

Besides the EW and mixed EW/QCD vertex corrections, one also has to
consider the impact of the unknown $\OO(\as^5)$ final-state QCD contribution.
Using again the assumption that the perturbative series approximately follows a
geometric series, one obtains for the total width 
\begin{equation}
\GZ: \quad \OO(\as^5) \sim \frac{\OO(\as^4)}{\OO(\as^3)}\OO(\as^4) \approx 0.04\mev. \label{errfsr}
\end{equation}
Other higher-order final-state QED and QCD effects, $e.\,g.$ of order
$\OO(\alpha\as^2)$ or $\OO(\alpha^2\as)$ are expected to be even smaller by the
same assessment method.

Combining eqs.~\eqref{errgz} and \eqref{errfsr} and the $\OO(\alpha^2_{\rm
bos})$ estimate in quadrature, the total theory error adds up to
$\delta\GZ\approx 0.5\mev$. 

%-------------------------------------------------------------
\begin{table}[tbp]
\renewcommand{\arraystretch}{1.2}
\begin{center}
\begin{tabular}{|l|l||l|l||l|r|}
\hline
$\Gamma_{e,\mu\,\tau}$ & 0.012~MeV &
$\Gamma_{u,c}$ & 0.12~MeV &
$R_\ell$ & $5\times 10^{-3}$ \\
$\Gamma_\nu$ & 0.014~MeV &
$\Gamma_b$ & 0.21~MeV &
$R_c$ & $5\times 10^{-5}$ \\
%\cline{3-4}
$\Gamma_{d,s}$ & 0.09~MeV &
$\GZ$ & 0.5~MeV &
$R_b$ & $1.5\times 10^{-4}$ \\
\hline
\end{tabular}
\end{center}
\vspace{-2ex}
\mycaption{Remaining theory uncertainty 
for the partial and total $Z$ widths and branching ratios, using the estimation
procedure described in the text.
\label{tab:err}}
\end{table}
%-------------------------------------------------------------

\vspace{\bigskipamount}
Applying the same procedure to the partial widths, one obtains the theory errors
listed in Table~\ref{tab:err}. For the ratios ($R_\ell$, $R_c$ and $R_b$), the
theory uncertainty has been simply estimated from the partial widths 
using Gaussian error propagation.

\vspace{\bigskipamount}
For the hadronic peak cross-section, the theory error can be evaluated from
$\sigma^0_{\rm had} \propto (\Gamma_e\Gamma_{\rm had}/\GZ^2)\,(1+\delta X)$.
In the first term, the impact of perturbative higher-order corrections partially
cancels in the ratio. As a result, the dominant uncertainty stems from the
$\delta X$ term, leading to the estimates
\begin{equation}
\begin{aligned}
&\sigma^0_{\rm had}: \; 
&\OO(\alpha^3) &\sim \sigma^0_{\rm had,Born}\delta X
\frac{\GZ^{(\alpha^2)}}{\GZ^{(\alpha)}} \sim 3.7 \text{ pb} , \\
&&\OO(\alpha^2\as) &\sim \sigma^0_{\rm had,Born}\delta X
\frac{\GZ^{(\alpha\as)}}{\GZ^{(\alpha)}} \sim 4.2 \text{ pb} ,
\end{aligned}
\label{errsig}
\end{equation}
where the total width, $\GZ$, has been used for the scaling on perturbative
orders, since both $\delta X$ and $\GZ$ are related to the imaginary part of the
$Z$ self-energy. The $\OO(\alpha^2_{\rm bos})$ contribution for $\sigma^0_{\rm
had}$ is estimated by squaring the bosonic one-loop corrections to the
partials widths, as above, and using Gaussian error propagation, resulting in an
error contribution of about 2~pb. The total theory error follows from combining
this with \eqref{errsig} in quadrature, yielding $\delta\sigma^0_{\rm had}
\approx 6$~pb.

%%%%%%%%%%%%%%%%%%%%%%%%%%%%%%%%%%%%%%%%%%%%%%%%%%%%%%%%%%%%%%

\section{Summary}
\label{sc:summ}

In this article, the full electroweak two-loop
corrections from diagrams with closed fermion loops to all partial widths of the
$Z$-boson within the Standard Model has been presented. Together with previous
results for the effective weak mixing angle \cite{swlept,swlept2,swbb} and the
hadronic $Z$-peak cross-section \cite{short}, this provides a complete
description of fermionic two-loop corrections to resonant $Z$-boson production
and decay at $e^+e^-$ colliders.  Precise predictions are given for the commonly
used experimental observables: the total width $\GZ$, the branching ratios
$R_\ell$, $R_c$ and $R_b$, and the hadronic peak cross-section $\sigma^0_{\rm
had}$. For convenient use by other researchers, simple parametrization formulas
are provided, which accurately reproduce the full result over large ranges of the
input parameters. 

Electroweak two-loop corrections to the total width, partial widths, branching
ratios, and peak cross-sections are sizeable and must be included in
phenomenological analyses of LEP1 and SLC data.
Compared to previous calculations, which use an expansion for large
values of the top-quark mass, the new results lead to moderately small shifts of
a few$\times 10^{-4}$ for these observables, thus giving confidence in the robustness of the perturbative expansion.

The added information from the full electroweak two-loop corrections helps to
estimate the intrinsic uncertainty from unknown higher-order corrections. The
theory error is found to be safely below the current experimental errors for all
$Z$-pole observables. However, additional work will be necessary to match the
precision of a future linear $e^+e^-$ collider \cite{ilcq}.

As a by-product, an updated result for the branching $R_b$ has been presented,
which improves on Ref.~\cite{rb} by including additional higher-order terms.

%%%%%%%%%%%%%%%%%%%%%%%%%%%%%%%%%%%%%%%%%%%%%%%%%%%%%%%%%%%%%%

\section*{Acknowledgments}

The author gratefully acknowledges many discussions and detailed
numerical comparisons with S.~Mishima, and feedback on the manuscript from
R.~Kogler.
This work has been supported in part by the National Science Foundation under
grant no.\ PHY-1212635.

%%%%%%%%%%%%%%%%%%%%%%%%%%%%%%%%%%%%%%%%%%%%%%%%%%%%%%%%%%%%%%

\appendix
\section{Final-state QED and QCD corrections}

The dominant contributions from final-state QED and QCD radiation can be captured
through factorizable radiator functions ${\cal R}_{\rm V,A}$ for the vector and
axial-vector part, respectively. They are known up to ${\cal O}(\as^4)$ for
massless final-state quarks and ${\cal O}(\as^3)$ for terms that depend on the
masses of the external quarks \cite{rad,rad2}. Additionally, the ${\cal O}(\alpha^2)$
contributions from diagrams with closed fermion loops \cite{Kataev:1992dg} are also included here.

Up to the precision required for this project, they read
\begin{align}
&\begin{aligned} 
{\cal R}_V(s) =\, &1 + \frac{3Q_f^2}{4}\,\frac{\alpha(s)}{\pi} + \frac{\as(s)}{\pi}
 - \frac{Q_f^2}{4}\,\frac{\alpha(s)}{\pi}\,\frac{\as(s)}{\pi}
 +Q_f^2\bigl [ C_{\gamma2} + 2C_2^\Pt(s/\mt^2) \bigr ] \Bigl (\frac{\alpha(s)}{\pi}\Bigr )^2 \\
&+\bigl [ C_{02} + C_2^\Pt(s/\mt^2) \bigr ] \Bigl (\frac{\as(s)}{\pi}\Bigr )^2
 + C_{03}\Bigl (\frac{\as(s)}{\pi}\Bigr )^3
 + C_{04}\Bigl (\frac{\as(s)}{\pi}\Bigr )^4 \\
&+ 12\frac{m_f^2}{s}\,\frac{\as(s)}{\pi}-6\frac{m_f^4}{s^2}\,,
\end{aligned}
\\[.5ex] 
&\begin{aligned} 
{\cal R}_A(s) =\, &1 + \frac{3Q_f^2}{4}\,\frac{\alpha(s)}{\pi} + \frac{\as(s)}{\pi}
 - \frac{Q_f^2}{4}\,\frac{\alpha(s)}{\pi}\,\frac{\as(s)}{\pi}
 +Q_f^2\bigl [ C_{\gamma2} + 2C_2^\Pt(s/\mt^2) \bigr ] \Bigl (\frac{\alpha(s)}{\pi}\Bigr )^2 \\
&+\bigl [ C_{02} + C_2^\Pt(s/\mt^2) \pm I_2(s/\mt^2)\bigr ] \Bigl (\frac{\as(s)}{\pi}\Bigr )^2
 +\bigl [ C_{03} \pm I_3(s/\mt^2)\bigr ] \Bigl (\frac{\as(s)}{\pi}\Bigr )^3 \\
&+\bigl [ C_{04} \pm I_4(s/\mt^2)\bigr ] \Bigl (\frac{\as(s)}{\pi}\Bigr )^4 
-6\frac{m_f^2}{s} - 22\frac{m_f^2}{s}\,\frac{\as(s)}{\pi}+6\frac{m_f^4}{s^2}\,,
\end{aligned}
\end{align}
where contributions of $\OO(m_f^6)$, $\OO(m_f^4\as)$, $\OO(m_f^2\as^2)$, and
$\OO(m_f^2\alpha)$ have been neglected.
For $f=e,\,\mu,\,\tau$ the terms with $\as$ vanish.
In the expressions above, $Q_f$ is the electric charge of the fermion $f$,
the $\pm$ sign applies to down/up-type fermions,
and
\begin{align}
C_{\gamma 2} &= -\frac{55}{6} + \frac{20}{3}\zeta_3, \\
C_{02} &= \frac{365}{24}-11\zeta_3 + \Bigl
 (-\frac{11}{12}+\frac{2}{3}\zeta_3\Bigr )n_q, \\
C_2^\Pt(x) &= x\Bigl (\frac{44}{675} - \frac{2}{135}\log x\Bigr ) + \OO(x^2), \\
C_{03} &= -6.63694 - 1.20013 n_q - 0.005178 n_q^2, \\
C_{04} &= -156.61 + 18.77 n_q - 0.7974 n_q^2 + 0.0215 n_q^3, \\
I_2(x) &= -\frac{37}{12} + \log x + \frac{7}{81}x + \frac{79}{6000}x^2 + \OO(x^3), \\
I_3(x) &= -15.9877 + \frac{67}{18} \log x + \frac{23}{12}\log^2 x + \OO(x), \\
I_4(x) &= 49.0309 - 17.6637\log x + 14.6597\log^2 x + 3.6736\log^3 x + \OO(x),
\end{align}
where $n_q=5$ is the number of light quarks.

In addition, there exists a singlet vector correction, which cannot be assigned
to individual partial widths, but only to the total hadronic $Z$ decay
\cite{rad,rad2}. It first enters at $\OO(\as^3)$ and is numerically very small,
so that it can be neglected for the purposes of this analysis.

%%%%%%%%%%%%%%%%%%%%%%%%%%%%%%%%%%%%%%%%%%%%%%%%%%%%%%%%%%%%%%

\section{Extended parametrization formulae}
\label{sc:fit2}

In section~\ref{sc:fit1}, the numerical results for the $Z$-boson partial
widths, branching ratio, and peak cross-section were presented in terms of a
simple parametrization formula, which provides an accurate description within
current allowed ranges for the SM input parameters. However, in global SM
fits the results may be needed over a larger range of input parameters.
For this purpose the following formula with additional coefficients is
introduced:
\begin{align}
&X = \begin{aligned}[t] 
&X_0 + a_1 L_\PH + a_2 L_\PH^2 + a_3 \Delta_\PH + a_4 \Delta_\PH^2
 + a_5 \Delta_\Pt + a_6 \Delta_\Pt^2 + a_7 \Delta_\Pt L_\PH + a_8
  \Delta_\Pt L_\PH^2 \\
 &+ a_9 \Delta_{\as} + a_{10} \Delta_{\as}^2 + a_{11} \Delta_{\as} L_\PH +
  a_{12} \Delta_{\as} \Delta_\Pt 
 + a_{13} \Delta_\alpha + a_{14} \Delta_\alpha L_\PH + a_{15} \Delta_\PZ, 
 \label{par2} \end{aligned} \\[1ex]
&L_\PH = \log\frac{\MH}{125.7\gev}, \quad
 \Delta_\PH = \frac{\MH}{125.7\gev}-1,\quad
 \Delta_\Pt = \Bigl (\frac{\mt}{173.2\gev}\Bigr )^2-1, \quad
 \nonumber \\
& \Delta_{\as} = \frac{\as(\MZ)}{0.1184}-1, \quad
 \Delta_\alpha = \frac{\Delta\alpha}{0.059}-1, \quad
 \Delta_\PZ = \frac{\MZ}{91.1876\gev}-1. \nonumber
\end{align}
Its range of validity is $70\gev < \MH < 1000\gev$, $165\gev < \mt < 190\gev$, 
$\as=0.1184\pm 0.0050$, $\Delta\alpha = 0.0590 \pm 0.0005$ and $\MZ = 91.1876
\pm 0.0084 \gev$, with the coefficients and maximal numerical deviations given in
Tab.~\ref{tab:fit2}.

%-------------------------------------------------------------
\begin{table}[p]
\renewcommand{\arraystretch}{1.3}
\small
\begin{tabular}{|l|rrrrrrrr|}
\hline
Observable & \multicolumn{1}{c}{$X_0$} & 
 \multicolumn{1}{c}{$a_1$} & \multicolumn{1}{c}{$a_2$} & 
 \multicolumn{1}{c}{$a_3$} & \multicolumn{1}{c}{$a_4$} & 
 \multicolumn{1}{c}{$a_5$} & \multicolumn{1}{c}{$a_6$} & 
 \multicolumn{1}{c|}{$a_7$} \\
\hline
$\Gamma_{e,\mu}$ [MeV] &
 83.966 & $-$0.1017 & $-$0.06352 & 0.05500 & $-$0.00145 & 0.8051 & $-$0.027 &
 $-$0.017\\
$\Gamma_{\tau}$ [MeV] &
 83.776 & $-$0.1016 & $-$0.06339 & 0.05488 & $-$0.00145 & 0.8036 & $-$0.026 &
 $-$0.017\\
$\Gamma_{\nu}$ [MeV] &
 167.157 & $-$0.1567 & $-$0.1194 & 0.1031 & $-$0.00269 & 1.258 & $-$0.13 & $-$0.020\\
$\Gamma_{u}$ [MeV] &
 299.936 & $-$0.5681 & $-$0.2636 & 0.2334 & $-$0.00592 & 4.057 & $-$0.50 & $-$0.058\\
$\Gamma_{c}$ [MeV] &
 299.859 & $-$0.5680 & $-$0.2635 & 0.2334 & $-$0.00592 & 4.056 & $-$0.50 & $-$0.058\\
$\Gamma_{d,s}$ [MeV] &
 382.770 & $-$0.6199 & $-$0.3182 & 0.2800 & $-$0.00711 & 3.810 & $-$0.25 & $-$0.060\\
$\Gamma_{b}$ [MeV] &
 375.723 & $-$0.5744 & $-$0.3074 & 0.2725 & $-$0.00703 & $-$2.292 & $-$0.027 & $-$0.013\\
$\GZ$ [MeV] & 
 2494.24 & $-$3.725 & $-$2.019 & 1.773 & $-$0.04554 & 19.63 & $-$2.0 & $-$0.36 \\
\hline
$R_\ell$ [$10^{-3}$] &
 20750.9 & $-$10.00 & $-$1.83 & 1.878 & $-$0.0343 & $-$38.8 & $-$11 & 1.2 \\
$R_c$ [$10^{-3}$] &
 172.23 & $-$0.034 & $-$0.0058 & 0.0054 & $-$0.00012 & 1.00 & $-$0.15 & $-$0.0074 \\
$R_b$ [$10^{-3}$] &
 215.80 & 0.036 & 0.0057 & $-$0.0044 & 6.2$\times$$10^{-5}$ & $-$2.98 & 0.20 & 0.020 \\
\hline
$\sigma^0_{\rm had}$ [pb] & 
 41488.4 & 3.88 & 0.829 & $-$0.911 & 0.0076 & 61.10 & 16 & $-$2.0 \\
\hline
\end{tabular}\\[1ex]
\begin{tabular}{|l|@{\hspace{.2em}}r@{\hspace{.7em}}rr@{\hspace{.7em}}rr@{\hspace{.7em}}rrr|c|}
\hline
Observable & \multicolumn{1}{c}{$a_8$} & \multicolumn{1}{c}{$a_9$} & 
 \multicolumn{1}{c}{$a_{10}$} & \multicolumn{1}{c}{$a_{11}$} & 
 \multicolumn{1}{c}{$a_{12}$} & \multicolumn{1}{c}{$a_{13}$} & 
 \multicolumn{1}{c}{$a_{14}$} & \multicolumn{1}{c|}{$a_{15}$} & max.\ dev. \\
\hline
$\Gamma_{e,\mu}$ [MeV] &
 0.0066 & $-$0.095 & $-$0.010 & $-$0.015 & 0.23 & $-$1.1 & 0.064 & 285 & $<0.0015$\\
$\Gamma_{\tau}$ [MeV] &
 0.0066 & $-$0.095 & $-$0.010 & $-$0.015 & 0.23 & $-$1.1 & 0.064 & 285 & $<0.0015$\\
$\Gamma_{\nu}$ [MeV] &
 0.0133 & $-$0.19 & $-$0.018 & $-$0.021 & 0.34 & $-$0.084 & 0.064 & 503 & $<0.002$\\
$\Gamma_{u}$ [MeV] &
 0.0352 & 14.26 & 1.6 & $-$0.081 & 1.7 & $-$11.1 & 0.19 & 1251 & $<0.006$\\
$\Gamma_{c}$ [MeV] &
 0.0352 & 14.26 & 1.6 & $-$0.081 & 1.7 & $-$11.1 & 0.19 & 1251 & $<0.006$\\
$\Gamma_{d,s}$ [MeV] &
 0.0420 & 10.20 & $-$2.4 & $-$0.083 & 0.65 & $-$10.1 & 0.19 & 1468 & $<0.006$\\
$\Gamma_{b}$ [MeV] &
 0.0428 & 10.53 & $-$2.4 & $-$0.088 & 1.2 & $-$10.1 & 0.19 & 1456 & $<0.006$\\
$\GZ$ [MeV] & 
 0.257 & 58.60 & $-$4.1 & $-$0.53 & 7.6 & $-$56.0 & 1.3 & 9256 & $<0.04$\\
\hline
$R_\ell$ [$10^{-3}$] &
 0.72 & 732.1 & $-$44  & $-$0.64 & 5.6 & $-$357 & $-$4.7 & 11771 & $<0.15$ \\
$R_c$ [$10^{-3}$] &
 0.00091 & 2.3 & 1.3 & $-$0.0013 & 0.35 & $-$1.2 & 0.014 & 37 & $<0.01$ \\
$R_b$ [$10^{-3}$] &
 $-$0.00036 & $-$1.3 & $-$0.84 & $-$0.0019 & 0.054 & 0.73 & $-$0.011 & $-$18 & $<0.01$ \\
\hline
$\sigma^0_{\rm had}$ [pb] & 
 $-$0.59 & $-$579.4 & 38 & $-$0.26 & 6.5 & 84 & 9.5 & $\!\!\!-$86152 & $<0.25$\\
\hline
\end{tabular}
\mycaption{Coefficients for the parametrization formula \eqref{par2} for various
observables ($X$). Within the ranges $70\gev < \MH < 1000\gev$, $165\gev < \mt < 190\gev$,
$\as=0.1184\pm 0.0050$, $\Delta\alpha = 0.0590 \pm 0.0005$ and $\MZ = 91.1876 \pm
0.0084 \gev$, the formula approximates the full result with maximal deviations
given in the last column.
\label{tab:fit2}}
\end{table}
%-------------------------------------------------------------

%%%%%%%%%%%%%%%%%%%%%%%%%%%%%%%%%%%%%%%%%%%%%%%%%%%%%%%%%%%%%%

\end{document}